\documentclass[,final,numberedheadings
  ]
  {aipproc}

\layoutstyle{6x9}

\begin{document}

\title{Statistical Mechanics of systems with long range interactions}


\author{David Mukamel}{
  address={Department of Physics of Complex Systems, The Weizmann Institute
  of Science, Rehovot 76100, Israel}
}

%\author{<author2>}{
%  address={<common address for author2 and author3>}
%}

%\author{<author3>}{
%  address={<common address for author2 and author3>}
%  ,altaddress={<author1 address>} % additional visiting address
%}

\begin{abstract}
 Recent theoretical studies of statistical mechanical properties of
 systems with long range interactions are briefly reviewed. In these
 systems the interaction potential decays with a rate slower than $1/r^d$ at
large distances $r$ in $d$ dimensions. As a result, these systems
are non-additive and they display unusual thermodynamic and
dynamical properties which are not present in systems with short
range interactions. In particular, the various statistical
mechanical ensembles are not equivalent and the microcanonical
specific heat may be negative. Long range interactions may also
result in breaking of ergodicity, making the maximal entropy state
inaccessible from some regions of phase space. In addition, in many
cases long range interactions result in slow relaxation processes,
with time scales which diverge in the thermodynamic limit. Various
models which have been found to exhibit these features are
discussed.

\end{abstract}

\maketitle

%%%%%%%%%%%%%%%%%%%%%%%%%%%%%%%%%%%%%%%%%%%%
%% MAINMATTER
%%%%%%%%%%%%%%%%%%%%%%%%%%%%%%%%%%%%%%%%%%%%

\section{Introduction}

  Long range forces are rather common in nature. These forces
  are typically derived from two body potentials which at large
  distance, $r$, decay as $1/r^s$ with $s \le d$ in $d$ dimensions.
  Examples include self gravitating systems $(s=1)$ \cite{Padmanabhan,Chavanis},
  dipolar ferroelectrics and ferromagnets $(s=3)$ \cite{Landau}, non-neutral
  plasmas $(s=1)$ \cite{Nicholson}, two dimensional geophysical vortices which
  interact via a weak, logarithmically decaying, potential $(s=0)$
  \cite{Chavanis}, charged particles interacting via their mutual
  electromagnetic fiels, such as in free electron laser \cite{Barre04}
  and many others (for recent reviews see \cite{LesHouches}).
  As a result of the long range nature of the
  interactions, these systems are non-additive, and the energy of
  homogeneously distributed particles in a volume $V$ scales
  super-linearly with the volume, as $V^{1+\sigma}$, with
  $\sigma=1-s/d \ge 0$. The lack of additivity leads to many
  unusual properties, both thermal and dynamical, which are not
  present in the more commonly studied systems with short range
  interactions. For example, as has first been pointed out by
  Antonov \cite{Antonov} and later elaborated by Lynden-Bell
  \cite{LyndenBell68,LyndenBell99}, Thirring \cite{Thirring70,Thirring71} and
  others, the entropy $S$ needs not be a concave function of the
  energy $E$, yielding negative specific heat within the
  microcanonical ensemble. Since specific heat is always positive
  when calculated within the canonical ensemble, this indicates
  that the two ensembles need not be equivalent. Recent studies
  have suggested the inequivalence of ensembles is particularly
  manifested whenever a model exhibits a first order transition
  within the canonical ensemble \cite{Barre01,Barre02}. Similar
  ensemble inequivalence between canonical and grand canonical
  ensembles has also been discussed \cite{Misawa}.

  Studies of the relaxation
  processes is systems with long range interactions in some models
  have shown that the relaxation of thermodynamically unstable
  states to the stable equilibrium state may be unusually slow, with
  a characteristic time which diverges with the number of particles,
  $N$, in the system \cite{Antoni95,Latora98,Latora99,Yamaguchi03,Yamaguchi04,Mukamel05}.
  This, too, is in contrast with relaxation
  processes in systems with short range interactions, in which the
  relaxation time does not scale with $N$. As a result long lived quasi-stationary
  states (QSS) have been observed in some models, which in the thermodynamic
  limit, do not relax to the equilibrium state. Non-additivity has
  been found to result, in many cases, in breaking of ergodicity.
  Here phase space is divided into disjoint domains separated by
  finite gaps in macroscopic quantities, such as the total
  magnetization in magnetic systems
  \cite{Mukamel05,Feldman98,Borgonovi04,Borgonovi06,Hahn05,Hahn06,Bouchet07}.
  Within local dynamics, these systems are thus trapped
  in one of the domains.

  Typically, the entropy, $S$, which is measured by the number
  of ways $N$ particles with total energy $E$ may be distributed
  in a volume $V$, scales linearly with the volume. This is
  irrespective of whether or not the interactions in the system are
  long ranged. On the other hand, in systems with long range
  interactions, the energy
  scales super-linearly with the volume. Thus, in the thermodynamic limit,
  the free energy $F=E-TS$ is dominated by the energy at any finite
  temperature $T$, suggesting that the entropy may be neglected altogether.
  This would result in trivial thermodynamics.
  However, in many real cases, when systems of finite size are
  considered, the temperature could be sufficiently high so that
  the entropic term in the free energy, $TS$, becomes comparable to
  the energy $E$. In such cases the entropy may not be neglected
  and the thermodynamics is non trivial. This is the case in some
  self gravitating systems such as globular clusters (see, for example
  \cite{Chavanis}). In order to
  theoretically study this limit, it is convenient to rescale the
  energy by a factor $V^{-\sigma}$ (or alternatively, to rescale the
  temperature by a factor $V^\sigma$), making the energy and the
  entropy contribution to the free energy of comparable magnitude.
  This is known as the Kac prescription \cite{Kac}. While systems described
  by this rescaled energy are extensive, they are non-additive in
  the sense that the energy of two isolated sub-systems is not equal
  to their total energy when they are combined together and are
  allowed to interact.

  A special case is that of dipolar ferromagnets, where the interaction
  scales as $1/r^3$ ($\sigma=0$). In this borderline case
  between long and short range interactions, the energy depends on
  the shape of the sample. It is well known that for ellipsoidal magnets,
  the contribution of the long distance part of the dipolar interaction leads to a
  mean-field type term in the energy. This results in an effective Hamiltonian
  $H \to H-DM^2/N$, where $M$ is the magnetization of the system
  and $D$ is a shape dependent coefficient known as the demagnetization factor.
  In this Hamiltonian, the long range interaction
  between dipoles becomes independent of their distance,
  making it particularly convenient for theoretical studies \cite{Campa07}.

  Non-additivity is a feature which is not limited to systems with long range
  interactions. In fact finite systems with short range
  interactions, in which surface and bulk energies are comparable,
  are also non-additive. Features such as negative specific heat in
  small systems (e.g. clusters of atoms) have been discussed a
  number of studies \cite{RMBell,Gross,Chomaz}.

  In the present paper we briefly review recent theoretical
  studies of systems with long range interactions where such
  properties have been explored. In Section (2) general
  considerations are presented, arguing for some of the unusual
  properties of systems with long range interactions. In Section (3)
  some features of canonical and microcanonical phase diagrams
  are discussed within a recently studied Ising model with
  mean-field type interaction. Ergodicity breaking is discussed in
  Section (4), and slow relaxation processes, as observed in a
  number of models, are discussed in Section (5). A summary and
  general outlook is finally given in Section (6).

\section{General considerations}

  We start by presenting some general considerations concerning
  thermodynamic properties of systems with long range
  interactions. In particular we argue that in addition to
  negative specific heat, or non-concave entropy curve, which
  could be realized in the microcanonical ensemble, this ensemble
  also yields discontinuity in temperature whenever a first order
  transition takes place.

\begin{figure}[ht]
%\begin{center}
%\includegraphics[width=6.00cm,height=4.00cm]{region.eps}
%%\includegraphics[height=.3\textheight]{non_concave.eps}
\includegraphics[height=.3\textheight]{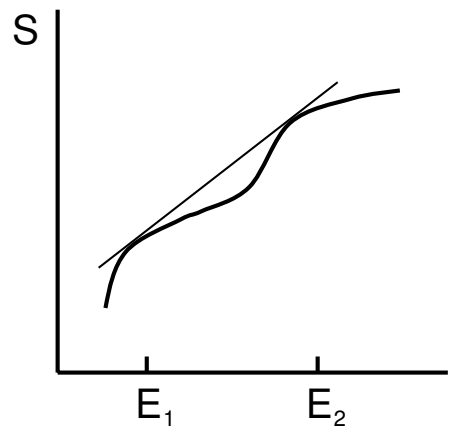}
\caption{\label{nonconcave_entropy} A non-concave entropy curve,
which for additive systems is made concave by the common tangent
line. In systems with long range interactions, the non-concave curve
may represent the actual entropy of the system, yielding negative
specific heat.}
%\end{center}
\end{figure}

  Consider the non-concave curve of Fig. (\ref{nonconcave_entropy}).
  For a system with short
  range interactions, this curve cannot represent the entropy
  $S(E)$. The reason is that due to additivity, the system
  represented by this curve is unstable in the energy interval
  $E_1 < E < E_2$. Entropy can be gained by phase separating the
  system into two subsystems corresponding to $E_1$ and $E_2$
  keeping the total energy fixed. The average energy
  and entropy densities in the coexistence region is given by the
  weighted average of the corresponding densities of the two
  coexisting systems. Thus the correct entropy curve in this
  region is given by the common tangent line,
  resulting in an overall concave curve. However, in systems with
  long range interactions, the average energy density of two
  coexisting subsystems is not given by the weighted average of the
  energy density of the two subsystems. Therefore, the non-concave
  curve of Fig. (\ref{nonconcave_entropy}) could, in principle,
  represent an entropy curve of a stable
  system, and phase separation need not take place. This results in negative
  specific heat. Since within the canonical ensemble specific heat
  is non-negative, the microcanonical and canonical ensembles are
  not equivalent. The above considerations suggest that the
  inequivalence of the two ensembles is particularly manifested
  whenever a coexistence of two phases is found within the
  canonical ensemble.

\begin{figure}[ht]
%\begin{center}
%\includegraphics[width=6.00cm,height=4.00cm]{region.eps}
\includegraphics[height=.3\textheight]{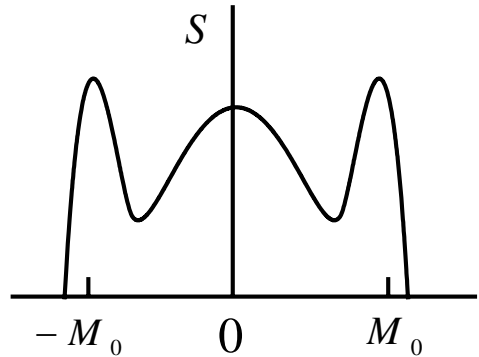}
\caption{\label{first_order} A typical entropy vs magnetization
curve of a magnetic system with long range interactions near a first
order transition at a given energy. As the energy varies the heights
of the peaks change and a first order transition is obtained at the
energy where the peaks are of equal height. }
%\end{center}
\end{figure}

\begin{figure}[ht]
%\begin{center}
%\includegraphics[width=6.00cm,height=4.00cm]{region.eps}
%%\includegraphics[height=.3\textheight]{first_order.eps}
\includegraphics[height=.3\textheight]{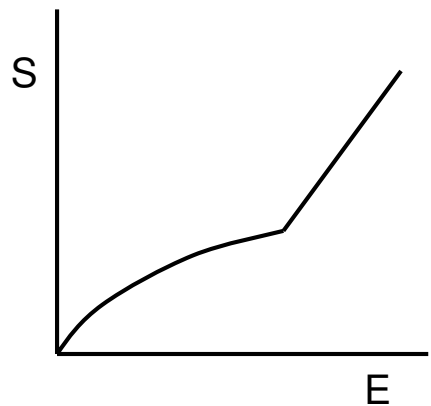}
\caption{\label{T_discontinuity} A typical entropy vs energy curve
for a system with long range interactions exhibiting a first order
transition. The slope discontinuity at the transition results in a
temperature discontinuity. }
%\end{center}
\end{figure}

  Another feature of systems with long range interactions is that
  within the microcanonical ensemble, first order phase transitions
  involve discontinuity of temperature. To demonstrate this point
  consider, for example, a magnetic system which undergoes a phase
  transition from a paramagnetic to a magnetically ordered phase.
  Let $M$ be the magnetization and $S(M,E)$ be the entropy of the
  system for a given magnetization and energy. A typical entropy
  vs magnetization curve for a given energy close to a first order
  transition is given in Fig. (\ref{first_order}). It exhibits
  three local maxima, one at $M=0$ and two other degenerate
  maxima at $\pm M_0$. At energies where the paramagnetic phase is
  stable, one has $S(0,E)>S(\pm M_0,E)$. In this phase the
  entropy is given by $S(0,E)$ and the temperature is obtained
  by $1/T= dS(0,E)/dE$. On the other hand at energies where the magnetic
  phase is stable, the entropy is given by $S(M_0,E)$ and the
  temperature is $1/T=dS(M_0,E)/dE$. At the first order transition
  point, where $S(0,E)=S(\pm M_0,E)$, the two derivatives are
  generically not equal, resulting in a temperature discontinuity.
  A typical entropy vs energy curve is given in Fig. (\ref{T_discontinuity}).

Systems with long range interactions are more likely to exhibit
breaking of ergodicity due to their non-additive nature. This may be
argued on rather general grounds. In systems with short range
interactions, the domain in the phase space of extensive
thermodynamic variables, such as energy, magnetization, volume etc.,
is convex. Let $\vec X$ be a vector whose components are the
extensive thermodynamic variables over which the systems is defined.
Suppose that there exist microscopic configurations corresponding to
two points $\vec X_1$ and $\vec X_2$ in this phase space. As a
result of the additivity property of systems with short range
interactions, there exist microscopic configurations corresponding
to any intermediate point between $\vec X_1$ and $\vec X_2$. Such
microscopic configurations may be constructed by combining two
appropriately weighted subsystems corresponding to $\vec X_1$ and
$\vec X_2$, making use of the fact that for sufficiently large
systems, surface terms do not contribute to bulk properties. Since
systems with long range interactions are non-additive, such
interpolation is not possible, and intermediate values of the
extensive variables are not necessarily accessible. As a result the
domain in the space of extensive variables over which a system is
defined needs not be convex. When there exists a gap in phase space
between two points corresponding to the same energy, local energy
conserving dynamics cannot take the system from one point to the
other and ergodicity is broken.

  These and other features of canonical and microcanonical phase
  diagrams are explored in the following sections by considering
  specific models.

\vspace{1.20cm}
\vspace{1.20cm}
\vspace{1.20cm}

  \section{Phase diagrams of models with long range interactions}
  \label{phase_diagrams}

  In order to obtain better insight into the thermodynamic behavior
  of systems with long range interactions it is instructive to
  analyze phase diagrams of representative models. A particularly
  convenient class of models is that where the long range part of
  the interaction is of mean-field type. In such models $\sigma=0$,
  and as pointed out above, they may be applied to study dipolar
  ferromagnets \cite{Campa07}. The insight obtained from studies of these models
  may, however, be relevant for other systems with $\sigma>0$, since
  the main feature of these models, namely non-additivity, is shares by
  all models with $\sigma \ge 0$.

  In recent studies both the canonical and microcanonical phase diagrams
  of some spin models with mean-field type long range interactions
  have been analyzed. Examples include discrete spin models such as the
  Blume-Emery-Griffiths model \cite{Barre01,Barre02} and the Ising model
  with long and short  range interactions \cite{Mukamel05}
  as well as continuous spin models of $XY$ type \cite{Buyl,Campa06}.
  These models are simple enough so that their thermodynamic
  properties can be evaluated in both ensembles. The common feature of these
  models is that their phase diagrams exhibit first and second order
  transition lines. In has been
  found that in all cases, the canonical and microcanonical
  phase diagrams differ from each other in the vicinity
  of the first order transition line. A classification of possible
  types of inequivalent canonical and microcanonical phase diagrams in
  systems with long range interactions is given in
  \cite{Bouchet05a}. In what follows we discuss in
  some detail the thermodynamics of one model, namely, the Ising model
  with long and short range interactions \cite{Mukamel05}.

  Consider an Ising model defined on a ring with $N$ sites. Let
  $S_i=\pm 1$ be the spin variable at site $i=1, \dots , N$. The
  Hamiltonian of the systems is composed of two interaction terms and is
  given by
\begin{equation}{\label{eq:Hamiltonian}}
H=-\frac{K}{2}{\sum_{i=1}^N}\left(S_iS_{i+1}-1\right)-\frac{J}{2N}
\left({\sum_{i=1}^N}S_i\right)^2.
\end{equation}
  The first term is a nearest neighbor coupling which could be either
  ferromagnetic $(K>0)$ or antiferromagnetic $(K<0)$. On the other
  hand the second term is ferromagnetic, $J>0$, and it corresponds to
  long range, mean-field type interaction. The reason for considering
  a ring geometry for the nearest neighbor coupling is that this is
  more convenient for carrying out the microcanonical analysis.
  Similar features are expected to take place in higher dimensions
  as well.

\begin{figure}[t]
%\begin{center}
\includegraphics[height=.3\textheight]{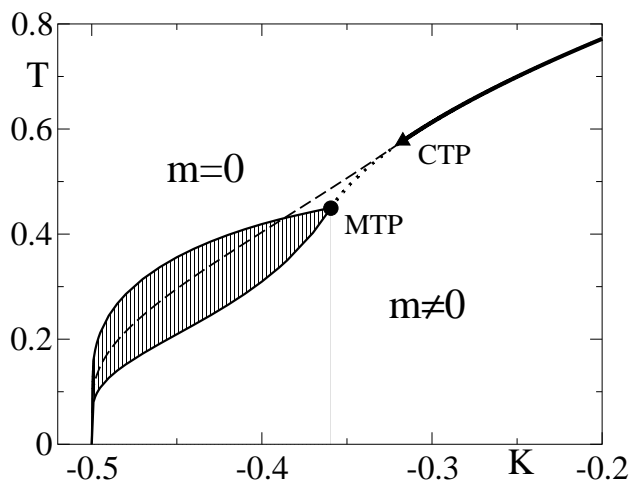}
%\includegraphics[width=6.00cm,height=4.00cm]{phase_diagram.eps}
%{phase_diagram_c.eps}%{phase_diagram_a.eps}
\caption{\label{fig:phase_diagram} The $(K,T)$ phase diagrams of the
model (\ref{eq:Hamiltonian}) within the canonical and microcanonical
ensembles. In the canonical ensemble the large $K$ transition is
continuous (bold solid line) down to the trictitical point CTP where
it becomes first order (dashed line). In the microcanonical ensemble
the continuous transition coincides with the canonical one at large
$K$ (bold line). It persists at lower K (dotted line) down to the
tricritical point MTP where it turns first order, with a branching
of the transition line (solid lines). The region between these two
lines (shaded area) is not accessible.}
%\end{center}
\end{figure}

  The canonical phase diagram of this model has been analyzed some
  time ago \cite{Nagle, Kardar}. The ground state of the model is ferromagnetic for
  $K>-J/2$ and is antiferromagnetic for $K<-J/2$. Since the system
  is one dimensional, and since the long range interaction term can
  only support ferromagnetic order, it is clear that for $K<-J/2$
  the system is disordered at any finite temperature, and no phase
  transition takes place. However, for $K>-J/2$ one expects
  ferromagnetic order at low temperatures. Thus a phase transition
  takes place at some temperature to a paramagnetic, disordered
  phase (see Fig. \ref{fig:phase_diagram}). For large $K$ the
  transition was found to be continuous, taking place at temperature
  given by
\begin{equation}
\label{critical_line}
  \beta=e^{-\beta K}.
\end{equation}
  Here $\beta = 1/T$, $J=1$ is assumed for simplicity, and $k_B=1$ is
  taken for the Boltzmann constant. The transition becomes first
  order for $K<K_{CTP}$, with a tricritical point located at an antiferromagnetic
  coupling $K_{CTP}=-\ln 3 /{2 \sqrt{3}}\simeq -0.317$. As usual, the
  first order line has to be evaluated numerically. The first order
  line intersects the $T=0$ axis at $K=-1/2$. The $(K,T)$ phase
  diagram is given in Fig. (\ref{fig:phase_diagram}).

  Let us now analyze the phase diagram of the model within the
  microcanonical ensemble \cite{Mukamel05}. To do this one has to calculate the
  entropy of the system for given magnetization and energy. Let
\begin{equation}
\label{U}
U=-\frac{1}{2}\sum_i\left(S_iS_{i+1}-1\right)
\end{equation}
  be the number of
  antiferromagnetic bonds in a given configuration characterized by
  $N_+$ up spins and $N_-$ down spins with $N_+ + N_- =N$. One would
  like to evaluate the number of microscopic configurations corresponding to
  $(N_+,N_-,U)$. Such configurations are composed of $U/2$ segments
  of up spins which alternate with the same number of segments of
  down spins, where the total number of up (down) spins is $N_+$
  $(N_-)$. The number of ways of dividing $N_+$ spins into $U/2$ groups
  is

  \begin{equation}
\left( \begin{array}{c} N_+ -1\\U/2-1
\end{array}\right )~,\\
\end{equation}
  with a similar expression for the down spins. To leading
  order in $N$, the number of configurations corresponding to
  $(N_+,N_-,U)$ is given by

\begin{equation}
\Omega(N_+,N_-,U) = \left( \begin{array}{c} N_+\\U/2
\end{array}\right )
\left( \begin{array}{c} N_-\\U/2
\end{array}\right )~.\\
\end{equation}
  Note that a multiplicative factor of order $N$ has been neglected in
  this expression, since only exponential terms in $N$ contribute to
  the entropy. This factor corresponds to the number of ways of
  placing the $U$ ordered segments on the lattice.
  Expressing $N_+$ and $N_-$ in terms of the number of spins, $N$,
  and the magnetization, $M=N_+ - N_-$, and denoting $m=M/N$,
  $u=U/N$ and the energy per spin $\epsilon=E/N$, one finds that
  the entropy per spin, $s(\epsilon,m)=\frac{1}{N}\ln\Omega$, is given
  in the thermodynamic limit by
\begin{eqnarray}{\label{eq:entropy}}
s(\epsilon,m)&=&\frac{1}{2}(1+m)\ln(1+m)+\frac{1}{2}(1-m)\ln(1-m)
\nonumber
\\&-& u\ln u -\frac{1}{2}(1+m-u)\ln(1+m-u)
\nonumber
\\& -&\frac{1}{2}(1-m-u)\ln(1-m-u)~,
\end{eqnarray}
  where $u$ satisfies
\begin{equation}
\label{energy} \epsilon=-\frac{J}{2}m^2+Ku ~\,.
\end{equation}
  By maximizing $s(\epsilon,m)$ with respect to $m$ one obtains both
  the spontaneous magnetization $m_s(\epsilon)$ and the entropy
  $s(\epsilon)\equiv s(\epsilon,m_s(\epsilon))$ of the
  system for a given energy $\epsilon$.

 In order to analyze the microcanonical phase transitions
 corresponding to this entropy we expand $s$ in powers of $m$,
\begin{equation}
s=s_0+Am^2+Bm^4 ~. \label{eq:Landau}
\end{equation}
Here the zero magnetization entropy is
\begin{equation}{\label{eq:s_0}}
s_0=-\frac{\epsilon}{K}\ln\frac{\epsilon}{K}-\left(1-
\frac{\epsilon}{K}\right)\ln\left(1-\frac{\epsilon}{K}\right)~,
%s_0=-\tilde\epsilon\ln\tilde\epsilon-\left(1-\tilde\epsilon\right)\ln\left(1-\tilde\epsilon\right),
%s_0=-\epsilon/K\ln\epsilon/K-(1-\epsilon/K)\ln(1-\epsilon/K)
\end{equation}
the coefficient $A$ is given by
\begin{equation}{\label{eq:A}}
A=\frac{1}{2}\left[\frac{1}{K}\ln\left(\frac{K-\epsilon}{\epsilon}\right)-\frac{\epsilon}{K-\epsilon}\right]~,
%A=\frac{1}{2}\left[\frac{\tilde\epsilon}{\tilde\epsilon-1}-\frac{1}{K}\ln\left(\frac{1-\tilde\epsilon}{\tilde\epsilon}\right)\right],
%A=-\frac{1}{2}\left[\frac{\tilde\epsilon}{1-\tilde\epsilon}-
%\frac{1}{K}\ln\left(\frac{\tilde\epsilon}{1-\tilde\epsilon}\right)\right],
\end{equation}
  and $B$ is another energy dependent coefficient which can be
  easily evaluated. In the paramagnetic phase both $A$ and $B$ are
  negative so that the $m=0$ state maximizes the entropy. At the
  energy where $A$ vanishes, a continuous transition to the
  magnetically ordered state takes place. Using the thermodynamic
  relation for the temperature
\begin{equation}
\frac{1}{T}=\frac{d s}{d\epsilon}~,
\end{equation}
  the caloric curve in the paramagnetic phase is found to be
\begin{equation}{\label{eq:T_null}}
\frac{1}{T}=\frac{1}{K}\ln\frac{K-\epsilon}{\epsilon}~.
\end{equation}
  This expression is also valid at the critical line where $m=0$.
  Therefore, the critical line in the $(K,T)$ plane may be evaluated
  by taking $A=0$ and using (\ref{eq:T_null}) to express $\epsilon$ in
  terms of $T$. One finds that the expression for the critical line is
  the same as that obtained within the canonical ensemble,
  (\ref{critical_line}).

  The transition is continuous as long as $B$ is negative, where the
  $m=0$ state maximizes the entropy. The transition changes its character
  at a microcanonical trictitical point where $B=0$. This takes
  place at $K_{MTP}\simeq -0.359$, which may be computed
  analytically using the expression for the coefficient $B$. The
  fact that $K_{MTP}<K_{CTP}$ means that while the microcanonical
  and canonical critical lines coincide up to $K_{CTP}$, the microcanonical
  line extends beyond this point into the region where,
  within the canonical ensemble, the model is magnetically ordered
  (see Fig. (\ref{fig:phase_diagram})). In this region the
  microcanonical specific heat is negative. For $K<K_{MTP}$
  the microcanonical transition becomes first order, and the
  transition line has to be evaluated numerically
  by maximizing the entropy. As discussed in the previous section,
  such a transition is characterized by temperature discontinuity.
  The shaded region in the $(K,T)$ phase diagram of Fig.
  (\ref{fig:phase_diagram}) indicates an inaccessible domain
  resulting from the temperature discontinuity.

  The main features of the phase diagram given in Fig.
  (\ref{fig:phase_diagram}) are not peculiar to the Ising
  model defined by the Hamiltonian (\ref{eq:Hamiltonian}), but are
  expected to be valid for any system in which a continuous line
  changes its character and becomes first order at a tricritical
  point. In particular, the lines of
  continuous transition are expected to be the same in both
  ensembles up to the canonical tricritical point. The
  microcanonical critical line extends beyond this point into the
  ordered region of the canonical phase diagram, yielding negative
  specific heat. When the microcanonical tricritical point is
  reached, the transition becomes first order, characterized by a
  discontinuity of the temperature. These features have been found in
  studies of other discrete spin models such as the spin-$1$
  Blume-Emery-Griffiths model \cite{Barre01,Barre02}. They have also
  been found in continuous spin models such as the $XY$ model with
  two- and four-spin mean-field like ferromagnetic interaction terms
  \cite{Buyl}, and in an $XY$ model with long and short range,
  mean-field type, interactions \cite{Campa06}.

\section{Ergodicity breaking}

Ergodicity breaking in models with long range interactions has
recently been explicitly demonstrated in a number of models such as
a class of anisotropic $XY$ models \cite{Borgonovi04,Borgonovi06},
discrete spin Ising models \cite{Mukamel05}, mean-field $\phi^4$
models \cite{Hahn05,Hahn06} and isotropic $XY$ models with four-spin
interactions \cite{Bouchet07}. Here we outline a demonstration of
this feature for the Ising model with long and short range
interactions defined in the previous section \cite{Mukamel05}.

\begin{figure}[ht]
%\begin{center}
%\includegraphics[width=6.00cm,height=4.00cm]{region.eps}
\includegraphics[height=.3\textheight]{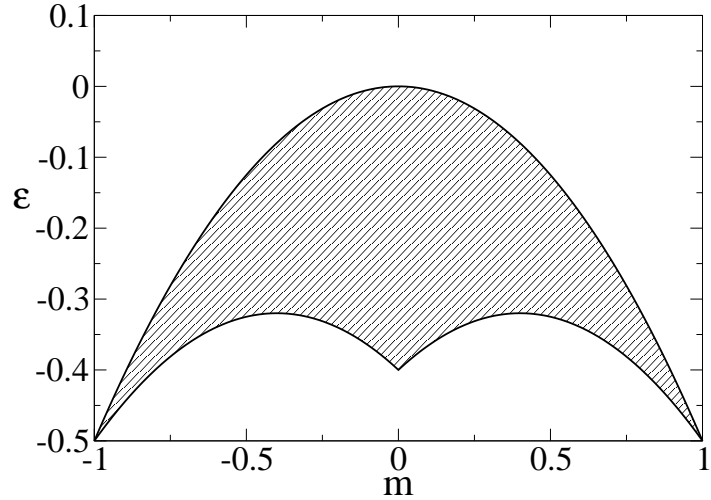}
\caption{\label{fig:Forbbiden_zone}Accessible region in the
$(m,\epsilon)$ plane (shaded area) of the Hamiltonian
(\ref{eq:Hamiltonian}) with $K=-0.4$. At low energies, the
accessible domain is composed of two disjoint magnetization
intervals and at intermediate energies three such intervals exist,
yielding ergodicity breaking. At higher energies the three intervals
join together and ergodicity is restored.}
%\end{center}
\end{figure}

Let us consider the Hamiltonian (\ref{eq:Hamiltonian}), and take,
for simplicity, a configuration of the spins with $N_+>N_-$. The
local energy $U$ is, by definition, non-negative. It also has an
upper bound which, for the case $N_+>N_-$, is $U \le 2N_-$. This
upper bound is achieved when the negative spins are isolated, each
contributing two negative bonds to the energy. Thus $0 \le u \le
1-m$. Combining this with (\ref{energy}) one finds that for positive
$m$ the accessible states have to satisfy
\begin{eqnarray}
\label{Mrestriction}
%%&&m\geq \sqrt{-2\epsilon} \quad ,\quad  m \geq m_+ \quad ,\quad  m \leq m_- \nonumber \\
&&m\leq \sqrt{-2\epsilon} \quad ,\quad  m \geq m_+ \quad ,\quad  m\leq m_- \nonumber \\
&&\mbox{with}\,\,  m_{\pm}=-K\pm \sqrt{K^2-2(\epsilon-K)}~.
\end{eqnarray}
Similar restrictions exist for negative $m$. These restrictions
yield the accessible magnetization domain shown in Fig.
(\ref{fig:Forbbiden_zone}) for $K=-0.4$.

The fact that the accessible magnetization domain is not convex
results in nonergodicity. At a given, sufficiently low energy, the
accessible magnetization domain is composed of two intervals with
large positive and large negative magnetization, respectively. Thus
starting from an initial condition which lies within one of these
intervals, local dynamics, to be discussed in the next section, is
unable to move the system to the other accessible interval, and
ergodicity is broken. At intermediate energy values another
accessible magnetization interval emerges near the $m=0$ state and
three disjoint magnetization intervals are available. When the
energy is increased the the three intervals join together and the
model becomes ergodic.
\vspace{1.20cm}

\section{Slow relaxation}

In systems with short range interactions the relaxation from a
thermodynamically unstable state is typically a fast process. For
example, in a magnetic, Ising like system, starting with a
magnetically disordered state at a low temperature, where the stable
state is the ordered one, the system will locally order in short
time. This leads to a domain structure in which the system is
divided into magnetically up and down domains of some typical size.
The domains forming process is fast in the sense that its
characteristic time does not scale with the system size. This domain
structure is formed by fluctuations, when a locally ordered region
reaches a critical size for which the loss its surface free energy
is compensated by the gain in its bulk free energy. This critical
size is independent of the system size, leading to a finite
relaxation time. Once the domain structure is formed it exhibits a
coarsening process in which the domains grow in size while their
number is reduced. This process, which is typically slow, eventually
leads to the ordered equilibrium state of the system.

This is very different from what happens in systems with long range
interactions. Here the initial relaxation from a thermodynamically
unstable state need not be fast and it could take place over a time
scale which diverges with the system size. The reason is that in the
case of long range interactions one cannot define a critical size of
an ordered domain, since the bulk and surface energies of a domain
are of the same order. It is thus of great interest to study
relaxation processes in systems with long range interactions and to
explore the types of behavior which might be encountered. In
principle the relaxation process may depend on the nature and
symmetry of the order parameter, say, whether it is discrete, Ising
like, or one with a continuous symmetry such as the $XY$ model. It
may also depend on the dynamical process, whether it is stochastic
or deterministic. In this section we briefly review some recent
results obtained in studies of the dynamics of some models with long
range interactions.

We start by considering the Ising model with long and short range
interactions defined in section (\ref{phase_diagrams}). The
relaxation processes in this model have recently been studied in
\cite{Mukamel05}. Since Ising models do not have intrinsic dynamics,
the common dynamics one uses in studying them is the Monte Carlo
(MC) dynamics, which simulates the stochastic coupling of the model
to a thermal bath. If one is interested in studying the dynamics of
an isolated system, one has to resort to the microcanonical MC
algorithm developed by Creutz ~\cite{Creutz} some time ago.
According to this algorithm a demon with energy $E_d\geq0$ is
allowed to exchange energy with the system. One starts with a system
with energy $E$ and a demon with energy $E_d=0$. The dynamics
proceeds by selecting a spin at random and attempting to flip it.
If, as a result of the flip, the energy of the system is reduced,
the flip is carried out and the excess energy is transferred to the
demon. On the other hand if the energy of the system increases as a
result of the attempted flip, the energy needed is taken from the
demon and the move is accepted. In case the demon does not have the
necessary energy the move is rejected. After sufficiently long time
and for large system size, $N$, the demon's energy will be
distributed according to the Boltzmann distribution
$\exp(-E_d/k_BT)$, where $T$ is the temperature of the system with
energy $E$. Thus, by measuring the energy distribution of the demon
one obtains the caloric curve of the system. Note that as long as
the entropy of the system is an increasing function of its energy,
the temperature is positive and the average energy of the demon is
finite. The demon's energy is thus negligibly small compared with
the energy of the system, which scales with its size. The energy of
the system at any given time is $E-E_D$, and it exhibits
fluctuations of finite width at energies just below $E$.

In applying the microcanonical MC dynamics to models with long range
interactions, one should note that the Boltzmann expression for the
energy distribution of the demon is valid only in the large $N$
limit. To next order in $N$ one has
\begin{equation}
P(E_D)\sim \exp{(-E_D/T-E_D^2/{2C_VT^2})}~,
\end{equation}
where $C_V=O(N)$ is the system's specific heat. In systems with
short range interactions, the specific heat is non-negative and thus
the next to leading term in the distribution function is a
stabilizing factor which may be neglected for large $N$. On the
other hand, in systems with long range interactions, $C_V$ may be
negative in some regions of the phase diagram, and on the face of
it, the next to leading term may destabilize the distribution
function. However the next to leading term is small, of order
$O(1/N)$, and it is straightforward to argue that as long as the
entropy is an increasing function of the energy, the next to leading
term does not destabilize the distribution. The Boltzmann
distribution for the energy of the demon is thus valid for large
$N$.

Using the microcanonical MC algorithm, the dynamics of the model
(\ref{eq:Hamiltonian}) has been studied in detail \cite{Mukamel05}.
Breaking of ergodicity in the region in the $(K,\epsilon)$ plane
where it is expected to take place has been observed.

The microcanonical MC dynamics has also been applied to study the
relaxation process of thermodynamically unstable states. It has been
found that starting with a zero magnetization state at energies
where this state is a local minimum of the entropy, the model
relaxes to the equilibrium, magnetically ordered, state on a time
scale which diverges with the system size as $\ln N$. The divergence
of the relaxation time is a direct result of the long range
interactions in the model.

The logarithmic divergence of the relaxation time may be understood
by considering the Langevin equation which corresponds to the
dynamical process. The equation for the magnetization $m$ is
\begin{equation}
\frac{\partial m}{\partial t}=\frac{\partial s}{\partial m} +\xi (t)
\,\,\,,\,\,\, <\xi(t) \xi(t')> = D \delta (t-t') \label{Langevin}
\end{equation}
where $\xi(t)$ is the usual white noise term. The diffusion constant
$D$ scales as $D \sim 1/N$. This can be easily seen by considering
the non-interacting case in which the magnetization evolves by pure
diffusion where the diffusion constant is known to scale in this
form. Since we are interested in the case of a thermodynamically
unstable $m=0$ state, which corresponds to a local minimum of the
entropy, we may, for simplicity, consider an entropy function of the
form
\begin{equation}
\label{toy_entropy}
 s(m)=am^2-bm^4
\end{equation}
with $a$ and $b$ non-negative parameters. In order to analyze the
the relaxation process we consider the corresponding Fokker-Planck
equation for the probability distribution $P(m,t)$ of the
magnetization at time $t$. It takes the form

\begin{equation}
\label{eq:FPE} \frac{\partial P(m,t)}{\partial t} =
D\frac{\partial^2 P(m,t)}{\partial m^2} -\frac{\partial}{\partial
m}\left(\frac{\partial s}{\partial m}P(m,t)\right)\;,
\end{equation}
This equation could be viewed as describing the motion of a particle
whose coordinate, $m$, carries out an overdamped motion in a
potential $-s(m)$ at temperature $T=D$. In order to probe the
relaxation process from the $m=0$ state it is sufficient to consider
the entropy (\ref{toy_entropy}) with $b=0$. With the initial
condition for the probability distribution $P(m,0)=\delta(m)$, the
large time asymptotic distribution is found to be \cite{Risken}
\begin{equation}
P(m,t) \sim \exp \left[ -\frac{ae^{-at}m^2}{D} \right] \,.
\end{equation}
This is a Gaussian distribution whose width grows with time. Thus,
the relaxation time from the unstable state, $\tau_{us}$, which
corresponds to the width reaching a value of $O(1)$, satisfies
\begin{equation}
\tau_{us} \sim -\ln D \sim \ln N~.
\end{equation}

The logarithmic divergence with $N$ of the relaxation time seems to
be independent of the nature of the dynamics. Similar behavior has
been found when the model (\ref{eq:Hamiltonian}) has been studied
within the Metropolis-type canonical dynamics at fixed temperature
\cite{Mukamel05}.

The relaxation  process from a metastable state (rather than an
unstable state discussed above) has been studied rather extensively
in the past. Here the entropy has a local maximum at $m=0$, while
the global maximum is obtained at some $m \ne 0$. As one would
naively expect, the relaxation time from the metastable $m=0$ state,
$\tau_{ms}$, is found to grow exponentially with $N$
\cite{Mukamel05}
\begin{equation}
\tau_{ms} \sim e^{N\Delta s}~.
\end{equation}
The entropy barrier corresponding to the non-magnetic state, $\Delta
s$, is the difference in entropy between that of the $m=0$ state and
the entropy at the local minimum separating it from the stable
equilibrium state. Such exponentially long relaxation times are
expected to take place independently of the nature of the order
parameter or the type of dynamics (whether it is stochastic or
deterministic). This has been found in the past in numerous studies
of canonical, Metropolis-type dynamics, of the Ising model with
mean-field interactions \cite{Griffiths}, in deterministic dynamics
of the $XY$ model \cite{Torcini} and in models of gravitational
systems \cite{Chavanis03}.

A different, rather intriguing, type of relaxation process has been
found in studies of the Hamiltonian dynamics of the $XY$ model with
mean-field interactions
\cite{Antoni95,Latora98,Latora99,Yamaguchi03,Yamaguchi04}. This
model has been termed the Hamiltonian Mean Field (HMF) model. In
this model, some non-equilibrium quasi-stationary states have been
identified, whose relaxation time grows as a power of the system
size, $N$, for some energy interval. This non-equilibrium stationary
state (which becomes a steady state in the thermodynamic limit)
exhibit some interesting properties such as anomalous diffusion
which have been extensively studied
(\cite{Latora99,Yamaguchi03,Yamaguchi04,Bouchet05}). At other energy
intervals the relaxation process has been found to be much faster,
with a relaxation time which grows as $\ln N$ \cite{Jain}. In what
follows we briefly outline the main results obtained for the HMF
model and for some generalizations of it.

The HMF model is defined on a lattice with each site occupied by an
$XY$ spin of unit length. The Hamiltonian takes the form
\begin{equation}
H=\sum_{i=1}^N \frac{p_{i}^{2}}{2}+\frac{1}{2N} \sum_{i,j=1}^N
\left[ 1-\cos(\theta_{i}-\theta_{j}) \right]~,
\end{equation}
where $\theta_i$ and $p_i$ are the phase and momentum of the $i$th
particle, respectively. In this model the interaction is mean-field
like. The model exhibits a continuous transition at a critical
energy $\epsilon_c=3/4$ from a paramagnetic state at high energies
to a ferromagnetic state at low energies. Within the Hamiltonian
dynamics, the equations of motion of the dynamical variables are
\begin{equation}
\frac{d \theta_i}{dt}= p_i \label{theta}~, \qquad  \qquad \frac{d
p_i}{dt}= -m_x \sin \theta_i+ m_y \cos \theta_i~, \label{mom}
\end{equation}
where $m_x$ and $m_y$ are the components of the magnetization
density
\begin{equation}
\vec{m}= \left(\frac{1}{N}\sum_{i=1}^{N} \cos \theta_i,
\frac{1}{N}\sum_{i=1}^{N} \sin \theta_i \right)~.
\end{equation}
The Hamiltonian dynamics obviously conserves both energy and
momentum. A typical initial configuration for the non-magnetic state
is taken as the one where the phase variables are uniformly and
independently distributed in the interval $\theta_i \in [-\pi,
\pi]$. A particularly interesting case is that where the initial
distribution of the momenta is uniform in an interval $[-p_0,p_0]$.
This has been termed the waterbag distribution. For such phase and
momentum distributions the initial energy density is given by
$\epsilon = p_0^2/6 - 1/2$.

Extensive numerical studies of the relaxation of the non-magnetic
state with the waterbag initial distribution have been carried out.
It has been found that at an energy interval just below $\epsilon_c$
this state is quasi-stationary, in the sense that the magnetization
fluctuates around its initial value for some time $\tau_{qs}$ before
it switches to the non-vanishing equilibrium value. This
characteristic time has been found to scale as
\cite{Yamaguchi03,Yamaguchi04}
\begin{equation}
\label{tauqs}
\tau_{qs} \sim N^\gamma
\end{equation}
with $\gamma \simeq 1.7$.

A very useful insight into the dynamics of the HMF model is provided
by analyzing the evolution of the probability distribution of the
phase and momentum variables, $f(\theta,p,t)$, within the Vlasov
equation approach \cite{Yamaguchi04}. It has been found that in the
energy interval $\epsilon^* < \epsilon < \epsilon_c$, with
$\epsilon^* = 7/12$, the waterbag distribution is linearly stable.
It is unstable for $\epsilon < \epsilon^*$. In this interval the
following growth law for the magnetization $m=\sqrt{m_x^2+m_y^2}~$
has been found \cite{Jain}:
\begin{equation}
m(t) \sim \frac{1}{\sqrt N} e^{\Omega t}~, \label{m_iso}
\end{equation}
where
\begin{equation}
\Omega=\sqrt{6(\epsilon^*-\epsilon)}~.
\end{equation}

The robustness of the quasi-stationary state to various
perturbations has been explored in a number of studies. The
anisotropic HMF model has recently been shown to exhibit similar
relaxation processes as the HMF model itself \cite{Jain}. The
anisotropic HMF model is defined by the Hamiltonian
\begin{equation}
H=\sum_{i=1}^N \frac{p_{i}^{2}}{2}+\frac{1}{2N} \sum_{i,j=1}^N
\left[ 1-\cos(\theta_{i}-\theta_{j})\right]- \frac{D}{2 N} \left[
\sum_{i=1}^N \cos \theta_{i} \right]^2 \label{H_an}~,
\end{equation}
where the anisotropy term with $D>0$ represents global coupling and
favors order along the $x$ direction. The model exhibits a
transition from  magnetically disordered to a magnetically ordered
state along the $x$ direction at a critical energy
$\epsilon_c=(3+D)/4$. An analysis of the Vlasov equation
corresponding to this model shows that as in the isotropic case, the
waterbag initial condition is stable for $\epsilon^* < \epsilon <
\epsilon_c$, where $\epsilon^*=(7+D)/12$. In this energy interval a
quasi-stationary state has been observed numerically, with a power
law behavior (\ref{tauqs}) of the relaxation time. The exponent
$\gamma$ does not seem to change with the anisotropy parameter.
Logarithmic growth in $N$ of the relaxation time is found for
$\epsilon < \epsilon^*$. A model with local, on site anisotropy term
has also been analyzed along the same lines \cite{Jain}. The model
is defined by the Hamiltonian
\begin{equation}
H=\frac{1}{2}\sum_{i=1}^{N}p_{i}^{2}+\frac{1}{2N}\sum_{i,j=1}^{N}(1-\cos(\theta_{i}-\theta_{j}))+
W \sum_{i=1}^{N}\cos^{2} \theta_{i}~. \label{H_on}
\end{equation}
Here, too, both types of behavior have been found.

Other extensions of the HMF model include the addition of short
range, nearest neighbor coupling to the Hamiltonian \cite{Campa06},
and coupling of the HMF model to a thermal bath, making the dynamics
stochastic \cite{Baldovin06}. In both cases quasi-stationarity is
observed with a power law growth of the relaxation time
(\ref{tauqs}) with an exponent $\gamma$ which seems to vary with the
interaction parameters of the models.

\section{Summary}
Some recent statistical mechanical studies of systems with long
range interactions have been reviewed. In these studies, various
properties of these systems, both thermal and dynamical, have been
explored within a class of models with mean-field type interactions.
Models with mean-field long range interaction are non-additive, and
as such they may be used to probe generic features of a wider class
of systems where non-additivity plays a major role, namely, systems
where the interaction between particles exhibits a power law decay
with their distance. The fact that some of the properties of
mean-field models may be exactly calculable, makes them particularly
interesting in this context.

The canonical and microcanonical phase diagrams of a number of
models have been calculated in recent studies. These include
discrete spin Ising like models, such as the spin-$1$
Blume-Emery-Griffiths model and the Ising model with long and short
range interactions, as well as continuous spin $XY$ like models with
either a fourth order global coupling or models with both long and
short range interactions. The common feature found in these studies
is that whenever the phase diagram exhibits a first order transition
the canonical and microcanonical ensembles become non-equivalent,
with the microcanonical ensemble exhibiting negative specific heat
and discontinuity in temperature. Recent studies comparing the grand
canonical and the canonical ensembles show that similarly, analogous
differences between these ensembles are present.

Rather general considerations indicate that systems with long range
interactions are likely to exhibit nonergodicity.  This is a direct
result of the fact that the domain in space of extensive variables,
such as energy, volume and magnetization, over which the model is
defined, is not necessarily convex.

Systems with long range interaction exhibit some intriguing
dynamical properties. Long relaxation times of thermodynamically
unstable states have been observed, with relaxation times diverging
with system size $N$. In some cases these times diverge with a power
law of $N$, in other cases the divergence is logarithmic with $N$.
In some models quasi-stationary states are found, which are long
lived non-equilibrium states which display unusual and intriguing
properties such as anomalous diffusion and algebraically long
relaxation times.

While studies of particular models provide useful insight as to the
possible properties of systems with long range interactions, an
overall scheme within which such properties can be classified is
still missing. For example the role of various parameters (such as
the symmetry and nature of the order parameter, the nature of the
dynamics-whether stochastic or deterministic etc.) in determining
the behavior of the system is not fully understood. It would be of
great interest to use the insight obtained from studies of specific
models in order to construct a more general guiding framework for
these systems.

\begin{theacknowledgments}
  I warmly thank J. Barr\'e, F. Bouchet, P. de Buyl, A. Campa, T.
  Dauxois, A. Giansanti, K. Jain, R. Khomeriki, S. Ruffo and N.
  Schreiber for fruitful and enjoyable collaboration on systems with
  long range interaction over the last several years. Support of the
  Minerva Foundation with funding from the Federal German Ministry
  for Education and Research, and of the Albert Einstein Center for
  Theoretical Research is gratefully acknowledged.
\end{theacknowledgments}

\end{document}